# NEW APPROACH OF ENVELOPE DYNAMIC ANALYSIS FOR MILLING PROCESS

CF. Bisu[1], M. Zapciu[1], A. Gérard[2], V. Vijelea[3], M. Anica[3]

[1] *University Politehnica of Bucharest, Laboratory Machines and Production Systems, Spl. Independentei, no.313. Bucharest, Romania.*
[2] *University Bordeaux 1, Mechanical and Physics Laboratory, Domaine Universitaire 33405 Talence Cedex- France.*
[3] *Digitline Company, Str. Baneasa, no. 2-6, sector 1, Bucharest, Romania.*

**Abstract:** This paper proposes a method to vibration analysis in order to on-line monitoring of milling process quality. Adapting envelope analysis to characterize the milling tool materials is an important contribution to the qualitative and quantitative characterization of milling capacity and a step by modeling the three-dimensional cutting process. An experimental protocol was designed and developed for the acquisition, processing and analyzing three-dimensional signal. The vibration envelope analysis is proposed to detect the cutting capacity of the tool with the optimization application of cutting parameters. The research is focused on FFT Fourier transform optimization of vibration analysis and vibration envelope to evaluate the dynamic behavior of the machine/ tool/workpiece.

## 1   INTRODUCTION

Increasing competitiveness and gaining a greater productivity and a better quality has been determined by increasing the performance of machine tools, of cutting tools and of designing and machining process. Given that these conditions the appearance of vibration is inevitable in the cutting process dynamic especially in the milling process. The vibrations analysis has long been used for the detection and identification of the machine tool condition. Also, the predictive maintenance is directed towards recognizing the earliest significant changes in machinery condition. Because of this, particular interest is the dynamic models of cutting in tridimensional conditions to provide stability information necessary to optimize existing processes work on the current market [Antonialli et al, 2010], [Engin et al, 2001]. This research is a part of a larger project of developing a dynamic three-dimensional cutting model by integrating the moments generated by cutting process [Cahuc et al, 2009], [Zapciu et al, 2009]. In a first step we are interested in analyzing the dynamic behavior of the mill tool and to obtain a complete method of monitoring the evolution of cutter teeth during the cutting process. By using modern acquisition systems and signal processing we are focused on the analysis of the contact tool/chip/workpiece so we can analyze the quality of work at any time of the cutter teeth.

## 2   OBJECTIVES

The method used in our research refers to the spectral envelope analysis to identify mechanical defects and obtaining a better response of the milling machine. Thus, the objective of the envelope analysis is providing real data on the milling capacity of the tool, tool wear and dynamic functionality of the assembly motor spindle with tool by emphasizing the dynamic behavior of the bearings.

## 3    EXPERIMENTAL SETUP

To achieve this research an experimental device is designed to obtain the dynamic information provided by the system: machine/tool/workpiece. The experiment were performed on a 3 axis CNC vertical machining center with 11 kW of power in the spindle motor and a maximum speed rotation of 8000 rpm. Wait for our goal the recording data of vibrations and cutting forces signals in the same time with rotational speed is absolutely necessary. A Kistler 9257B stationary dynamometer Quartz 3- Component, a National Instruments NI USB-6216 analogical/digital data acquisition board and Fastview software were used for three axis cutting force measurements. A three-dimensional PCB piezoelectric accelerometer fixed on the workpiece and a B&K unidirectional piezoelectric accelerometer placed on the spindle in radial direction, a National Instruments NI USB-9162 analogical/digital board and Fastview software were used for vibrations measurement.

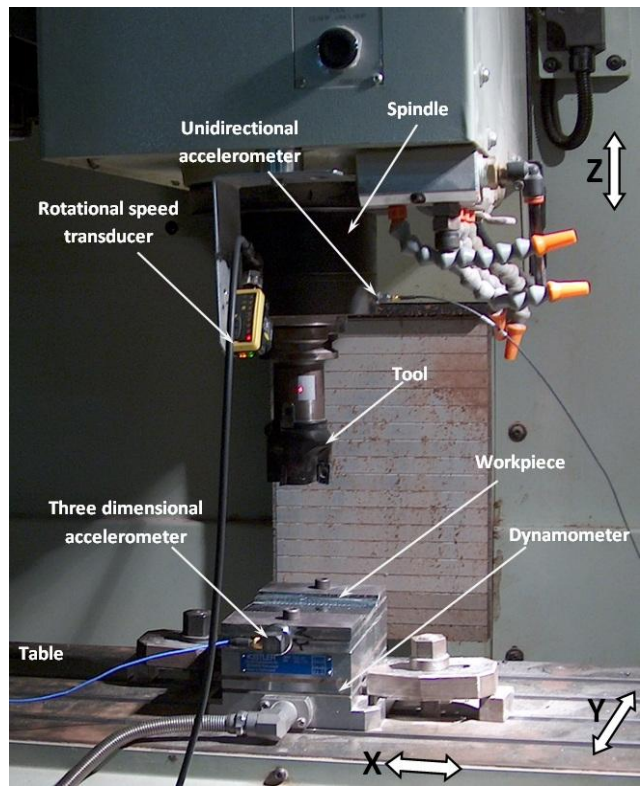

*Figure 1; Experimental setup.*

The speed of rotation is achieved through a laser sensor tachometer. The signals were processed with Fastview program, application developed with Digitline Company. Before starting the dynamic analysis of the tool in cutting process, the characterization of the machine is necessary in order to identify the dynamics of the assembly spindle – bearings. An experimental protocol has been established in order to conduct a thorough analysis of the spindle in different conditions **[**Bisu et al., 2010**]**. The cutting tool was also analyzed by impact vibrations using a PCB piezoelectric three-dimensional accelerometer and an instrumented hammer with a B&K force transducer in order to identify their transfer function in a broad range of frequencies. Samples were recorded at 25 kHz. Figure 2 show the FFT signal measured in three-directions after the impact hammer impulse. Test were performed on a steel materials E24-2 (RSt 37-2), the tool milling cutter were used here is R365-080Q27-S15M265259 with 80 mm diameter and 6 teeth recommended by tool manufacturer for face milling with a 0.1 mm/dents for feed and 340 m/min for cutting speed.

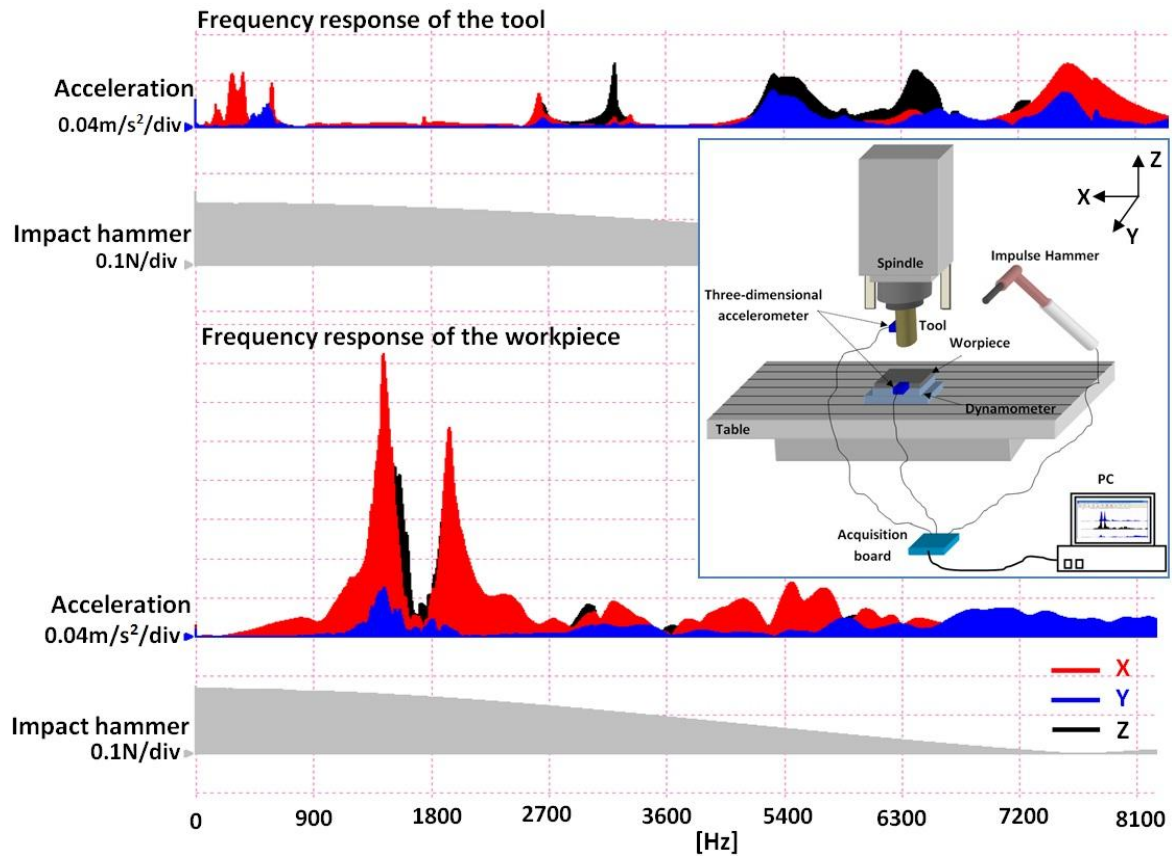

*Figure 2; Hammer impact test setup for the tool and workpiece.*

## 4 THEORY AND SIMULATION

This paper analyzes the load on each tooth of the milling tool synchronization of input/output while cutting tooth with tool rotation speed and angular position determination of each tooth. Also is made to highlight the different dynamic phenomena such as imbalance, non-alignment or eccentricity.

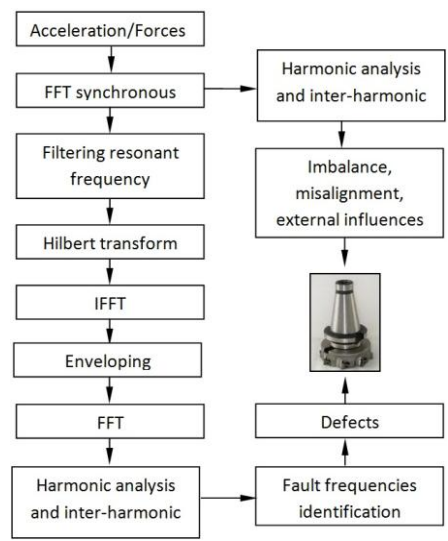

*Figure 3; Identification of the defects using envelope analysis.*

The tool chosen for calculation and simulation is a mill cutter with 6 teeth. The developed calculation has the purpose to determine by simulation the evolution of the teeth in machining process. For these aspects we used the envelope analysis method (figure 3), adapted to the case of milling process, respectively the dynamic analysis of the milling tool. To understand the behavior of the milling cutter we have analyzed two cases of the tool: when the teeth are perfectly identical, and when the teeth are asymmetrical. And so in the first case we obtained by enveloping method the existence of a single frequency pick while positioned on *(N/60)x6*, where N is rotational speed (rpm). In the second case, with asymmetrical teeth, we got along main frequency *(N/60)x6* the existence of a lower frequency with a equivalent amplitude to the main frequency. The further step is the cutting test with the aim to apply this method in a real case.

## 5     RESULTS AND ANALYSIS

The study is focused on dynamic behavior analysis of the mill cutter during the cutting process with 0.5 mm and 1 mm depth of cut. Before applying the enveloping method requires the identification of dynamic frequencies that occur in the cutting process.

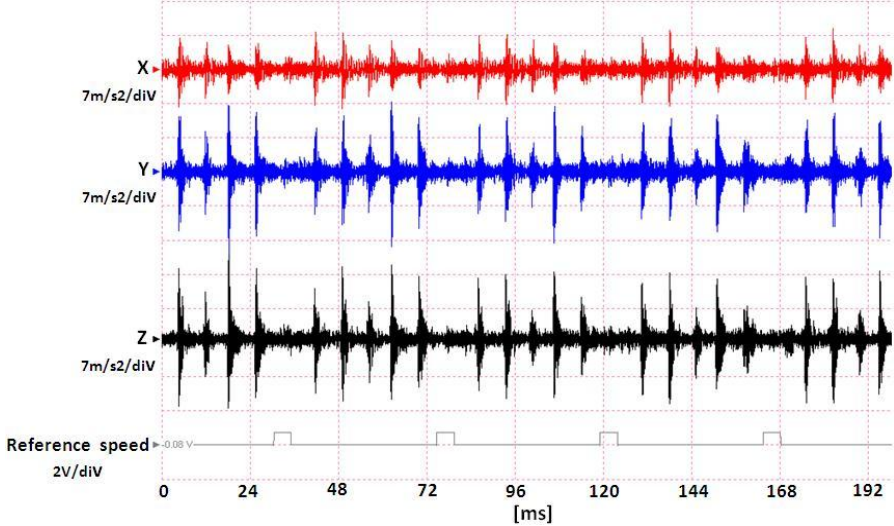

*Figure 4; Vibration cutting signal for x, y and z direction for 0.5 mm depth of cut.*

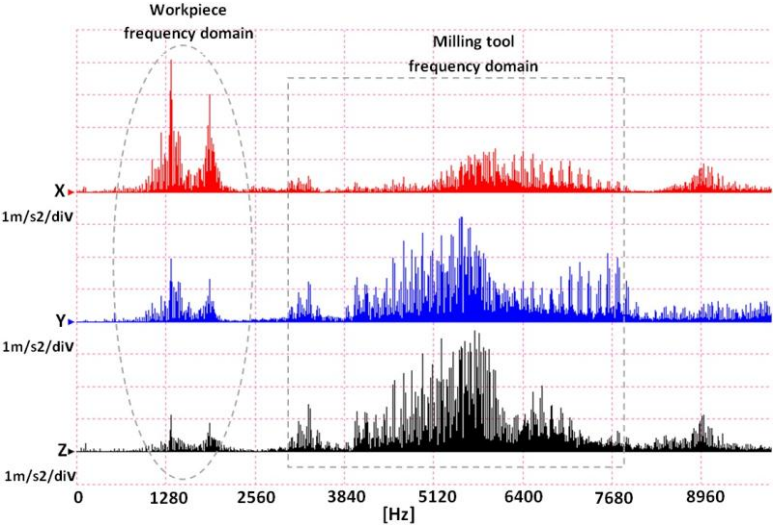

*Figure 5; Frequency spectrum during the cutting process.*

Thus in figure 4 is presented the cutting signal on the three-directions to which we apply FFT transform, represented in figure 5. Knowing the excitation frequency of the tool and the workpiece we can identify the dynamic influences of each of them in the spectrum. With dynamic information tool and workpiece can apply the enveloping synchronous method to evaluate with high precision the cutting quality of the tool (figure 6). Also we can see that in the figure 6 one of the cutter teeth do not have a proper load. The objective here is to accurately track the transfer power by tooth workpiece contact and its harmonic distribution.

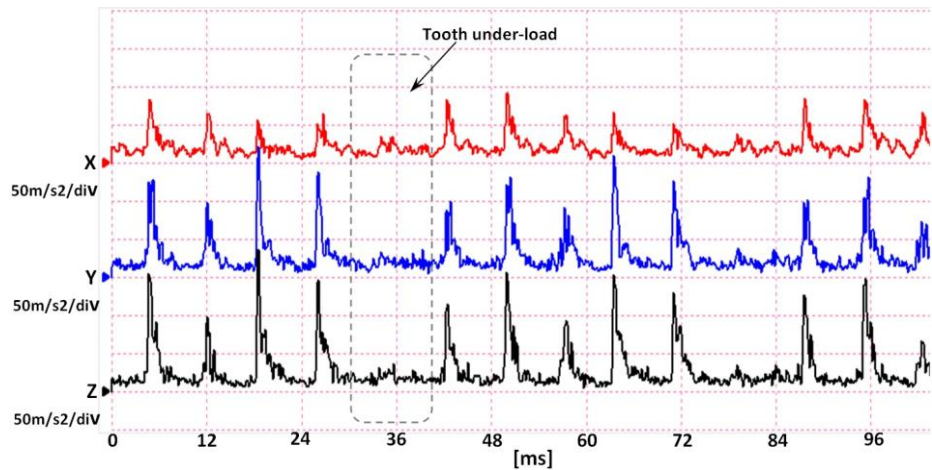

*Figure 6; Envelope vibration signal for the three-directions.*

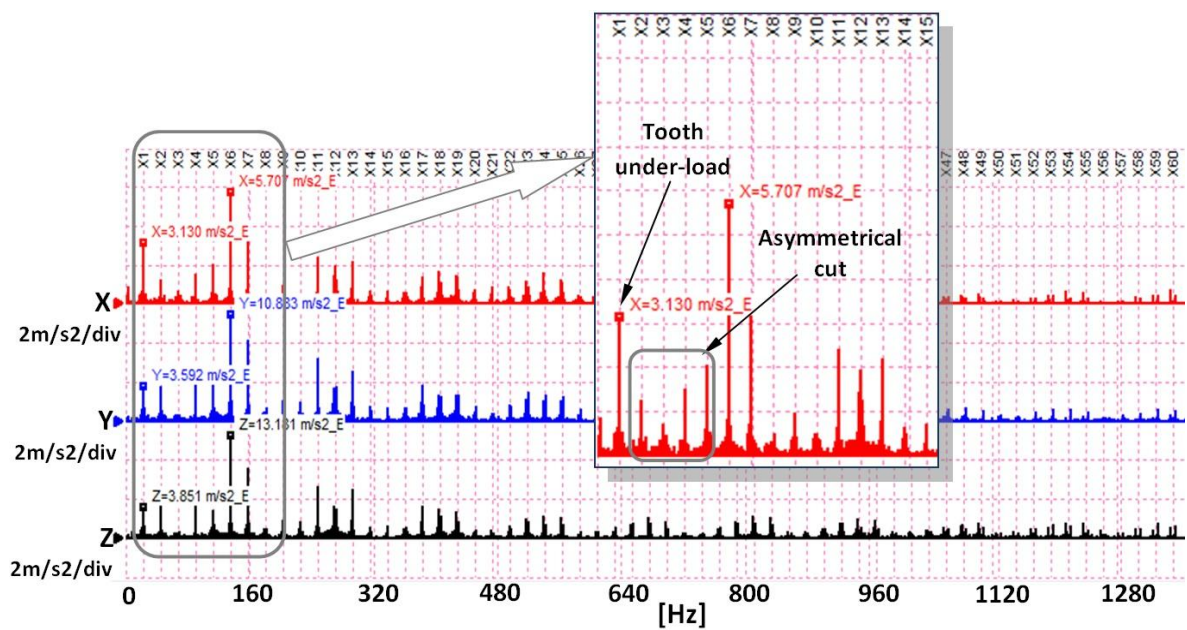

*Figure 7; The result of synchronous FFT analysis of vibration envelope.*

The aim of the envelope method applied to the milling process is achieved by frequency domain processing, consistent in high accuracy synchronous FFT transform, filtering resonance band of workpiece and tool, Hilbert transform [Kalvoda et al, 2010] followed by Inverse Fast Fourier Transform (IFFT), figure 6. Next the FFT analyses of the envelope ensure high precision description of the mill cutter to identify the type and amplitude of asymmetry and wear (figure 7). Each cutter tooth asymmetry is automatically qualified through the harmonic components with a lower frequency than the principal frequency equivalent of teeth number. The same method is applied in the case of forces, highlighting the quality of cut and obtained the similarly results with the vibrations (figure 8).

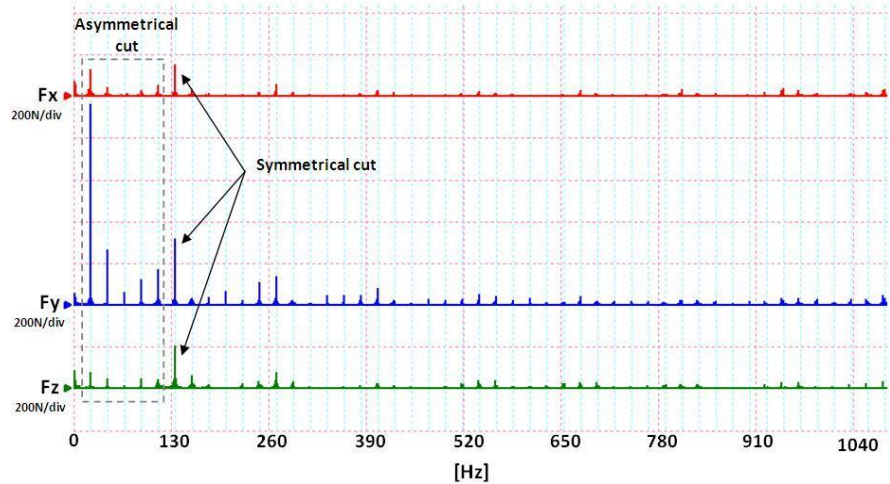

*Figure 8; Frequency spectrum of the three-direction forces components.*

## 6   CONCLUSION

This paper proposed a method of dynamic analysis based on envelope analysis method with the purpose to identify and evaluate the dynamic behavior of the tool during cutting process. An experimental protocol was designed and developed for the acquisition, processing and analyzing of the three-dimensional vibration and force signal. This method is useful both, dynamic characterization of the tool and also for the monitoring process and maintenance. In the future we are interested in creating a dynamic three-dimensional model useful in optimizing the milling process which takes into account these results.

**Acknowledgement**: This paper was supported by CNCSIS-UEFISCSU, project PNII-RU-code-194/2010.

E-mail address: cfbisu@gmail.com
Tel.: +40-7-24 01 62 95; fax: +40-2-14 02 97 24